# A structural model of quarks and leptons.

Clemens HEUSON
Zugspitzstr. 4 , D-87493 Lauben , Germany
email: clemens.heuson@freenet.de

A model is proposed in which quarks, leptons and gauge bosons are composites of magnetically charged preons (rishons) $T(q=1/3)$ and $V(q=0)$ with magnetic charges $g=g_0(1,2,-3)$. Structural formulas of composite particles and their interactions are given, with the binding lines representing magnetic fields. The three quark colors are generated by different magnetic charges of the constituents. Three higher generations are produced by adding rishon-antirishon pairs with the three magnetic charges.

## 1. Introduction

It is unlikely that the experimentally extremely successful standard model of particle physics is a fundamental theory, since it contains a large number of particles and unexplained parameters. One way to reduce the number of states and parameters is the assumption that quarks, leptons and gauge bosons are composites of more fundamental particles called preons. We will repeat some arguments supporting this view:

- There are simply too many quarks, leptons and gauge bosons, together at least 60 particles and most likely some more, compare this with the about 110 composite elements of the periodic table.
- There are about 30 unexplained parameters like masses, coupling constants and mixing parameters including neutrinos, which may be derived from a theory with fewer fundamental states.
- The charge connection between quarks and leptons $\Sigma q_i = 0$ in one generation as required by anomaly cancellation can be explained by common constituents.
- The family pattern with higher generation fermions differing only in mass can be understood in terms of extra neutral constituents in crude analogy to isotopic nuclei.
- The instability of higher generation fermions and weak bosons disqualifies them as fundamental particles, since historically all unstable systems have been composite.
- Changes of particle identities in interactions (except annihilation/creation) may imply that the involved particles are composite in analogy to chemical reactions understood as regrouping of constituents.
- Compositeness is the way, which nature has chosen from galaxies down to quarks and leptons.

Several composite models have been proposed, see for example the review of Lyons [1]. The original rishon model [2] has been introduced as an attractive and economical counting scheme reproducing the first generation of quarks and leptons. The main drawback was the lack of an dynamical theory and today presumable nobody believes, that the three quark colors can be obtained by ordering of only two rishons. Later on a dynamical scheme was proposed, in which a $SU(3)_H$ hypercolor-force binds rishons with appropriate $SU(3)_C$ colors [3]. We find it not very attractive to assume an SU(3) like force with just another strength of coupling, because on the way down from molecules to quarks and leptons nature never repeated itself in structure and interactions. On the other hand one has then a large number of preons and (hyper-) gluons, so that the original goal of reducing the number of states cannot be achieved. As an alternative way for preon binding magnetic forces have been suggested [4]. Wu [5] proposed a variation of the rishon model with electric- and magnetic-like charges, a drawback was however that the simple addition of electric charges $e/3$ and 0 reproducing the first generation has been lost. Koh, Pati, Rodrigues as mentioned in [5] suggested a rishon model with magnetic charges $(1,3,-4)g_0$ excluding exotic states, but now the question arises, why nature should have omitted the magnetic charge $2g_0$. In this letter we propose a model of magnetically charged rishons with the charge assignment $g=(1,2,-3)g_0$. While this does not sound very new in the first moment, we shall see that one gets a nice structural representation of composite quarks, leptons, gauge bosons and their interactions in analogy to chemical formula of molecules with the binding lines representing magnetic fields instead of electron pairs. A simple picture of the color and generation structure of composite fermions and their interactions with four sequential generations of quarks and leptons is provided.

## 2. The preons

As in the original rishon model we assume that the quarks and leptons of the first generation are composed of two fundamental rishons $T(q=1/3)$, $V(q=0)$ and the corresponding antirishons $\overline{T}, \overline{V}$ with spin 1/2. We assume furthermore that rishons carry magnetic charges. From electroweak interactions of leptons and the asymptotic freedom of quarks one knows, that composite fermions don't carry magnetic charges. For this reason the nonzero magnetic charges $g = g_0(i,j,k)$ of the three rishons in a quark or lepton must cancel to zero:
$i + j + k = 0$
From the magnetic charge quantisation [6] with n an integer

$$q_1 g_2 - q_2 g_1 = \frac{\hbar c}{2} n$$

applied to the system $T_i V_j$ one obtains $\frac{e}{3} g_j = \frac{\hbar c}{2} n$ or $g_j = \frac{3\hbar c}{2e} n$. Similarly for other systems this gives

$$g = n g_0 \text{ with } g_0 = \frac{3\hbar c}{2e}$$

Since the magnetic charges must cancel to zero, several charge assignments are possible. The most simple ones are: $g/g_0 = (1,1,-2), (1,2,-3), (1,3,-4)$. The first one cannot explain the three quark colors, and the third rises the question why nature should have omitted the simple magnetic charge of size $g=2g_0$. So we choose the second assignment providing us with the following fundamental spin 1/2 preons and antipreons:

$$T_1, T_2, T_{-3}, V_1, V_2, V_{-3}, \overline{T}_{-1}, \overline{T}_{-2}, \overline{T}_3, \overline{V}_{-1}, \overline{V}_{-2}, \overline{V}_3$$

T-rishons are electrically and magnetically charged dyons, while V-rishons are magnetically charged monopoles. The states carry the magnetic charge in multiples of $g_0$ as index. The electric charges are $q(T) = -q(\overline{T}) = e/3$ and $q(V) = q(\overline{V}) = 0$. Now one would like to know the masses of these fundamental preons, which cannot be answered in the affirmative. One possibility is that they are all massless, avoiding the introduction of further mass parameters. Alternatively one may assume that rishons with different magnetic and electric charges carry different masses, as would be suggested by the classical dyon mass formula
$m = a\sqrt{q^2 + g^2}$ [7].

## 3. Composite fermions

From the above magnetic charge assignment $g = g_0(1,2,-3)$ it is clear that composite quarks and leptons of the first generation have the structures $R_1 - R_{-3} = R_2$ or $\overline{R}_{-1} - \overline{R}_3 = \overline{R}_{-2}$ where R=T,V. The single (double) line represents an attractive magnetic binding between a monopole/dyon and an antimonopole/antidyon of strength one (two). One may also imagine that the lines resemble magnetic strings connecting the magnetically charged rishons. Of course one has to consider composite states as dynamical rather than rigid structures. Due to quantum field theory one expects a sea of virtual particles within any composite state. We now write down all composite quarks and leptons of the first generation.

| | | | |
|---|---|---|---|
| $e^-$: | $\overline{T}_{-1} - \overline{T}_3 = \overline{T}_{-2}$ | $e^+$: | $T_1 - T_{-3} = T_2$ |
| $u_R$: | $V_1 - T_{-3} = T_2$ | $\overline{u}_R$: | $\overline{V}_{-1} - \overline{T}_3 = \overline{T}_{-2}$ |
| $u_B$: | $T_1 - V_{-3} = T_2$ | $\overline{u}_B$: | $\overline{T}_{-1} - \overline{V}_3 = \overline{T}_{-2}$ |
| $u_G$: | $T_1 - T_{-3} = V_2$ | $\overline{u}_G$: | $\overline{T}_{-1} - \overline{T}_3 = \overline{V}_{-2}$ |
| $d_R$: | $\overline{T}_{-1} - \overline{V}_3 = \overline{V}_{-2}$ | $\overline{d}_R$: | $T_1 - V_{-3} = V_2$ |
| $d_B$: | $\overline{V}_{-1} - \overline{T}_3 = \overline{V}_{-2}$ | $\overline{d}_B$: | $V_1 - T_{-3} = V_2$ |
| $d_G$: | $\overline{V}_{-1} - \overline{V}_3 = \overline{T}_{-2}$ | $\overline{d}_G$: | $V_1 - V_{-3} = T_2$ |
| $\nu_e$: | $V_1 - V_{-3} = V_2$ | $\overline{\nu}_e$: | $\overline{V}_{-1} - \overline{V}_3 = \overline{V}_{-2}$ |

One sees that one has three quarks states arbitrarily assigned to the three quark colors R,G,B and exactly one state for each lepton. The mysterious ordering of rishons in a quark [2] now has a simple explanation: the three quark colors are distinguished by the magnetic charge of the rishon differing from the other two. At first sight it seems natural that all these states have comparable masses because of their similar structure. We can't however provide any dynamical calculations, since the three body problem is notoriously difficult, as is known from the case of baryons which are three quark states. Therefore here and in the following we cannot answer questions concerning the masses of composite states. For similar reasons we cannot say, why spin 3/2 fermions are not observed, they may be excluded on dynamical reasons or have higher masses. One may speculate that the strong binding between $R_{-3}$ and $R_2$ produces a scalar dipreon, so that composite states have spin 1/2 as in the model in [8]. Another problem mentioned in [5] also arises here: the three color states are not absolutely degenerate because of uneven electric charge interactions. This effect should however be small because of the large magnetic binding. Finally because of the present magnetic charge assignment, contrary to the (1,3,-4) case, exotic composite fermions may be constructed as $R_1 - \overline{R}_{-2} - R_1$ and $\overline{R}_{-1} - R_2 - \overline{R}_{-1}$. As will be clear from the structure of composite bosons in section 4, these exotic states cannot couple to composite weak bosons and gluons, so if existing they could belong to dark matter.

The existence of three generations I: $\nu_e, e^-, u, d$  II: $\nu_\mu, \mu^-, c, s$  III: $\nu_\tau, \tau^-, t, b$ with the same quantum numbers as the first one but with different approximately exponentially increasing masses is an unsolved puzzle generalising the old electron-muon puzzle (remember Rabi's comment to the muon: who ordered that?). It is well known that radiative transitions between different generations are not observed, which beside the wrong mass differences compared to the compositeness scale presumably exclude radial excitations. The simplest solution is obtained by adding rishon-antirishon-pairs to the first generation. Since we have rishons of magnetic charges $|g|=g_0(1,2,3)$ at our disposal, we can construct three higher generations by adding three $R_i \overline{R}_i$ - pairs for i=1,2,-3. Since only three sequential generations have been observed up to now, fourth generation fermions and especially the fourth neutrino must be very heavy, to escape the experimental limit from the $Z^0$-width. Because of the unknown rishon masses and the lack of dynamical calculations one cannot say which pair corresponds to which generation. Two effects seem to work against each other. If rishons with higher $|g|$ have higher masses they should correspond to higher generations. One the other side rishons with greater magnetic charge should exhibit a stronger magnetic binding and therefore the resulting states should have lower mass. Due to the different structures it is conceivable, that generations differ in mass. We write down one possibility, bearing in mind, that the reverse order of generations II-IV may be true.

$$
\begin{array}{llll}
& \quad R_1 \quad \overline{R}_{-1} & \quad R_2 - \overline{R}_{-2} & \quad \overline{R}_3 = R_{-3} \\
& \quad | \quad\quad | & \quad | \quad\quad | & \quad | \quad\quad | \\
f_I: \; R_1 - R_{-3} = R_2 \quad f_{II}: \; R_1 - R_{-3} - R_2 & f_{III}: \; R_1 - R_{-3} - R_2 & f_{IV}: \; R_1 - R_{-3} - R_2
\end{array}
$$

The problem arising here is that it is not clear a priori if the additional pairs consist of $T\overline{T}$ or $V\overline{V}$ - pairs or eventually mixtures of them. Certainly this has consequences for quark mixture and neutrino oscillations.

## 4. Composite gauge bosons

Here we consider the possibility that some or perhaps all gauge bosons of the standard model are composites of the fundamental rishons. Beginning with the weak bosons $W^\pm$ and $Z^0$ it seems likely that they are six-rishon-states. The charged weak bosons are $W^+ = TTTVVV$ and $W^- = \overline{TTTVVV}$ in order to enable processes like $e^- + W^+ \to \nu_e$ , $\overline{u} + W^+ \to \overline{d}$. The neutral weak boson is $Z^0 = RRR\overline{RRR}$ to enable $f + Z^0 \to f$, where f is any quark or lepton and R = T,V. The structure formulas are:

$$
\begin{array}{ll}
W^+: \; T_1 - T_{-3} - T_2 & Z^0: \; R_1 - R_{-3} - R_2 \\
\quad\quad\; | \quad\quad | & \quad\quad\; | \quad\quad | \\
\quad\quad V_2 - V_{-3} - V_1 & \quad\quad \overline{R}_{-1} - \overline{R}_3 - \overline{R}_{-2}
\end{array}
$$

Other isomeric states of the W - boson may be obtained by interchanging the positions of $T_i$ and $V_i$ for i=1, 2, -3. Interestingly these three states can mediate the processes $\overline{u}_C + W^+ \to \overline{d}_C$ where C=R,G,B are the three quark-colors, while the displayed state gives $e^- + W^+ \to \nu_e$. One understands that W-bosons can decay in weakly interacting leptons as well as in quarks with color and how 6-rishon states can form, which is not clear in the models in [2], [3] or [5].

Since quark colors are explained by different magnetic charges of the rishons and gluons carry the same quantum numbers as a quark-antiquark-pair, it seems inevitable that the gluons $G_\alpha$ ($\alpha$=1..8) are composites mediating processes like $u_R \rightarrow u_B$, $u_B \rightarrow u_G$, $u_G \rightarrow u_R$ and similar ones for the d-quarks. The gluons therefore must have the following structure:

$$G_{\overline{RB}}: \quad \overline{V}_{-1} - \overline{T}_3 - \overline{T}_{-2} \qquad G_{\overline{BG}}: \quad \overline{T}_{-1} - \overline{V}_3 - \overline{T}_{-2} \qquad G_{\overline{GR}}: \quad \overline{T}_{-1} - \overline{T}_3 - \overline{V}_{-2}$$
$$\quad\quad\quad\; | \quad\quad | \qquad\qquad\qquad\qquad\; | \quad\quad | \qquad\qquad\qquad\qquad\; | \quad\quad |$$
$$\quad\quad\quad T_1 - V_{-3} - T_2 \qquad\qquad\qquad T_1 - T_{-3} - V_2 \qquad\qquad\qquad V_1 - T_{-3} - T_2$$

The analogous gluons for the d-quarks are obtained by replacing $T_i \overline{T}_i \leftrightarrow V_i \overline{V}_i$ where i=2,1,-3 in the above order. This structural representation could give an explanation for the box-structure of gluons introduced by Shupe [2] and how 'ordering' of rishons in a quark is changed by gluons. Because of the 6-rishon-structure of the gluons similar to W and Z, it is unclear why their mass should be zero as predicted by QCD ? Also unanswered remains the question of quark confinement.

While it is entirely possible that the photon $\gamma$ is elementary, we'll consider the option of it's compositeness. Since the photon couples only to charged particles it should consist of $T_i \overline{T}_i$ -pairs. One could envisage the photon as six-rishon state more in agreement with unification or as two rishon-state [2]. There are arguments in favour of the last case: a photon must be able to couple to a d-quark containing one charged rishon, so only one $T_i \overline{T}_i$ -pair with parallel spins (↑↑) can be involved in this interaction, and because we assume that T-rishons carry electric charge e/3 the electric interaction should also exist at the level of preons. Therefore a composite photon should be a mixture of the structures:

$$\gamma: \quad T_1 - \overline{T}_{-1}, T_2 = \overline{T}_{-2}, T_{-3} \equiv \overline{T}_3$$

The above picture of a composite photon of course faces several problems like the zero photon mass, photon selfinteractions or the possible existence of $\overline{V}V$ - states.

Finally we consider a possible compositeness of the graviton g with spin 2 interacting with all particles without changing flavor or color [9]. If composite it would in the simplest case be a mixture of $g = RR\overline{RR}$ states (R=T,V) with parallel spins (↑↑↑↑). As in the case of the photon, the question why the graviton should be massless cannot be answered. Several structures for the graviton are possible and presumable the graviton is a mixture of them. Both rishon-antirishon-pairs should be involved in interactions giving the following structure formulas:

$$g: \qquad R_1 - \overline{R}_{-2} \qquad\qquad R_2 - R_{-3} \qquad\qquad R_1 - R_{-3}$$
$$\qquad\qquad\quad | \qquad\qquad\qquad\quad | \quad\;\; \| \qquad\qquad\qquad\quad \|$$
$$\qquad\quad \overline{R}_{-1} - R_2 \qquad\qquad \overline{R}_{-2} - \overline{R}_{-3} \qquad\qquad \overline{R}_{-1} - \overline{R}_3$$

The different coupling strengths of the four fundamental interactions between quarks and leptons should in principle be derivable from these structures.

The standard model contains one Higgs boson, responsible for the particle masses, while in most of its extensions there are several Higgs bosons. In general scalar states can be constructed as spin zero 2,4 or 6 rishon states and it is difficult to say without any experimental information which of them correspond to the observed states.

## 5. Composite particle interactions

Weak interactions with charged bosons $W^\pm$ change the flavor of the involved particles, while those involving the neutral $Z^0$ preserve flavor. As examples we display the following interactions with structural formulas:

$$u_R + W^- \to d_R \qquad e^+ + W^- \to \bar{\nu}_e \qquad \nu_e + Z^0 \to \nu_e$$

$$\begin{array}{ccc}
V_1 & \overline{V}_{-1} & \overline{V}_{-2} \\
| & | & || \\
T_{-3} & + \overline{T}_3 & - \overline{V}_{-2} \to \overline{V}_3 \\
|| & | & | \\
T_2 & \overline{T}_{-2} & - \overline{V}_3 \\
 & | & \\
 & \overline{T}_{-1} &
\end{array}
\qquad
\begin{array}{cccc}
T_1 & \overline{T}_{-1} & \overline{V}_{-2} & \\
| & | & | & \\
T_{-3} & + \overline{T}_3 & - \overline{V}_{-2} & \to \overline{V}_3 \\
|| & | & | & || \\
T_2 & \overline{T}_{-2} & - \overline{V}_3 & \overline{T}_{-1} \\
\end{array}
\qquad
\begin{array}{cccc}
V_1 & \overline{V}_{-1} & V_1 & V_1 \\
| & | & | & | \\
V_{-3} & + \overline{V}_3 & - V_{-3} & \to V_{-3} \\
|| & | & | & || \\
V_2 & \overline{V}_{-2} & - V_2 & V_2 \\
 & | & & \\
 & \overline{V}_{-1} & &
\end{array}$$

Three incoming rishons are annihilated by three antirishons of the weak boson $W^-$, while the remaining three form the outcoming state. From the first reaction one sees that $W^-$ changes flavor but not the color of quarks, which in the original rishon model was mysteriously ascribed to an amorphous structure. Quarkmixing in charged weak interactions can occur via $R_i \overline{R}_i \to R_j \overline{R}_j$ or energy and the transition amplitudes should in principle determine the elements of the CKM-matrix.

Strong interactions are described by gluons changing the color of quarks. We display the interaction of some gluons with an u and d-quark:

$$u_R + G_{\overline{R}B} \to u_B \qquad d_R + G_{\overline{R}B} \to d_B \qquad u_R + G_{\overline{R}R} \to u_R$$

$$\begin{array}{cccc}
V_1 & \overline{V}_{-1} & T_1 & T_1 \\
| & | & | & | \\
T_{-3} & + \overline{T}_3 & - V_{-3} & \to V_{-3} \\
|| & | & | & || \\
T_2 & \overline{T}_{-2} & - T_2 & T_2
\end{array}
\qquad
\begin{array}{cccc}
\overline{T}_{-1} & T_1 & \overline{V}_{-1} & \overline{V}_{-1} \\
| & | & | & | \\
\overline{V}_3 & + V_{-3} & - \overline{T}_3 & \to \overline{T}_3 \\
|| & | & | & || \\
\overline{V}_{-2} & V_2 & - \overline{V}_{-2} & \overline{V}_{-2}
\end{array}
\qquad
\begin{array}{cccc}
V_1 & \overline{V}_{-1} & V_1 & V_1 \\
| & | & | & | \\
T_{-3} & + \overline{T}_3 & - T_{-3} & \to T_{-3} \\
|| & | & | & || \\
T_2 & \overline{T}_{-2} & - T_2 & T_2
\end{array}$$

The gluon coupling $d_R$ and $d_B$ is obtained from the gluon coupling $u_R$ and $u_B$ by replacing $T_2 \overline{T}_{-2} \leftrightarrow V_2 \overline{V}_{-2}$.

The coupling of a gluon to a lepton is not allowed, since only two of the six rishons constituting an gluon could annihilate with the corresponding antirishons of the lepton. So the present picture can explain why gluons couple only to quarks but not to leptons and change color but not flavor. In a similar manner gluons can couple to quarks of higher generations.

The coupling of a photon to a charged quark or lepton is shown in the cases

$$u_R + \gamma \to u_R \qquad d_R + \gamma \to d_R \qquad e^+ + \gamma \to e^+$$

$$\begin{array}{ccc}
V_1 & T_{-3} & V_1 \\
| & ||| & | \\
T_{-3} & + \overline{T}_3 & \to T_{-3} \\
|| & & || \\
T_2 & & T_2
\end{array}
\qquad
\begin{array}{ccc}
\overline{T}_{-1} & T_1 & \overline{T}_{-1} \\
| & | & | \\
\overline{V}_3 & + \overline{T}_{-1} & \to \overline{V}_3 \\
|| & & || \\
\overline{V}_{-2} & & \overline{V}_{-2}
\end{array}
\qquad
\begin{array}{ccc}
T_1 & & T_1 \\
| & & | \\
T_{-3} & + T_2 & \to T_{-3} \\
|| & || & || \\
T_2 & \overline{T}_{-2} & T_2
\end{array}$$

One rishon is annihilated by the corresponding antirishon and flavor or color are not changed. The so constructed photon can couple to all charged particles.

A composite graviton couples to all composite particles without changing flavor or color. Two rishons of the fermion are annihilated by the corresponding two antirishons of the graviton and the fermion remains unchanged in structure. One example is:

$$u_R + g \to u_R$$

$$
\begin{array}{ccccc}
V_1 & & & & V_1 \\
| & & & & | \\
T_{-3} + \overline{T}_3 & = & T_{-3} & \to & T_{-3} \\
\| & | & | & & \| \\
T_2 & \overline{T}_{-2} & - T_2 & & T_2
\end{array}
$$

With the exception of the photon, if constructed as two-rishon-state, all composite gauge bosons can couple only to composite fermions. The pleasant feature of this approach is that we have an explanation, why the composite fermions have the observed interactions. The unpleasant feature is that we cannot explain the masses of the gauge bosons and the different coupling constants.

Several exotic bosons can be constructed and we show as examples a leptoquark X, as needed for proton decay, changing a lepton in a quark and a colored W-boson changing flavor and color of quarks.

$$e^+ + X^{-1/3} \to u_R \qquad\qquad d_R + W^+_{RB} \to u_B$$

$$
\begin{array}{cccc}
T_1 & \overline{T}_{-1} & V_1 & V_1 \\
| & | & | & | \\
T_{-3} + \overline{T}_3 & - T_{-3} & \to & T_{-3} \\
\| & | & | & \| \\
T_2 & \overline{T}_{-2} & - T_2 & T_2
\end{array}
\qquad
\begin{array}{cccc}
\overline{T}_{-1} & T_1 & & T_2 \\
| & | & & \| \\
\overline{V}_3 + V_{-3} & - T_2 & \to & V_{-3} \\
\| & | & | & | \\
\overline{V}_{-2} & V_2 & - V_{-3} & T_1 \\
& & | & \\
& & T_1 &
\end{array}
$$

Indirect CP-violation requires the transition $\overline{K}^0(s\overline{d}) \to K^0(d\overline{s})$, which in the standard model is described through a box diagram involving W-bosons and up-like quarks. In our structural model this process is effectively described as an exchange, where the $R_1\overline{R}_{-1}$ - pair moves from the $s$-quark to the $\overline{d}$-quark giving a $d$-quark and a $\overline{s}$-quark. A similar thing would happen in the transition $\overline{B}^0(b\overline{d}) \to B^0(d\overline{b})$ with the $R_2\overline{R}_{-2}$ - pair.

Neutrino oscillations are transitions $\nu_i \to \nu_j$ in vacuum or matter, which in our model can be described as transitions $R_i\overline{R}_i \to R_j\overline{R}_j$ or energy.

## 6. Summary

In summary we have proposed a structural model of composite quarks and leptons built by two rishons $T(q=1/3)$ and $V(q=0)$ each with magnetic charges $g=g_0(1,2,-3)$ together with their antiparticles. The model preserves much of the simplicity of the original rishon model. One understands why composite fermions have the observed interactions with their properties and gains an explanation of higher generations by adding rishon-antirishon pairs with different magnetic charges and of the color of quarks in terms of different magnetic charges of the rishons. The main problem is the lack of dynamical calculations, so that many questions concerning particle masses, coupling strengths and mixing parameters cannot be answered. A severe theoretical problem is of course: can nonabelian quantum field theories of weak and strong interactions be obtained from a model of this kind? However since in many composite models weak bosons are considered as composite, having an effective nonabelian theory may not be so strange as seems in the first moment.

Addendum: A possible fourth sequential generation of heavy quarks and leptons does not contradict recent electroweak precision data, see [10].